# A Blind Robust Watermarking Scheme Based on SVD and Circulant matrices

Noui Oussama[1] and Noui Lemnouar[2]

[1]Department of Computer science, Faculty of Science University of Batna
`os.noui@gmail.com`
[2] Department of Mathematics, Faculty of Science University of Batna
`nouilem@yahoo.fr`

## ABSTRACT

*Multimedia security has been the aim point of considerable research activity because of its wide application area. The major technology to achieve copyright protection, content authentication, access control and multimedia security is watermarking which is the process of embedding data into a multimedia element such as image or audio, this embedded data can later be extracted from, or detected in the embedded element for different purposes. In this work, a blind watermarking algorithm based on SVD and circulant matrices has been presented. Every circulant matrix is associated with a matrix for which the SVD decomposition coincides with the spectral decomposition. This leads to improve the Chandra algorithm [1], our presentation will include a discussion on the data hiding capacity, watermark transparency and robustness against a wide range of common image processing attacks.*

## KEYWORDS

*Digital image watermarking, Singular value Decomposition, circulant matrix, ownership protection.*

## 1. INTRODUCTION

Due to improvements in the digital image technology and growing availability and usability of internet during the past several years, demands for storage and transmitting of digital images have seen a distinct increase, Unfortunately, the problem of illegal piracy is increasingly serious. Protection of digital multimedia content has become an increasingly important issue for content owners and service providers. Encryption data was a way to ensure only the owner to view the content, there are still ways for illegal using of the content after decryption [2, 3], that lead us to a new method for protection. Watermarking is the process of embedding data called a watermark into the multimedia object such that watermark can be detected or extracted later to make an assertion about the object. The object may be an audio, image or video and even 3-D models [4, 5]. Watermarking algorithms fall into two categories. The first form of watermarking was a spatial watermarking technique work with the pixel values directly. Generally, spatial domain watermarking is easy to implement from a computational point of view, but too fragile to resist numerous attacks [6, 7]. In spatial domain, the watermark is directly embedded into the specific pixels of the host image, but in transform domain which our proposed scheme based in the watermark is embedded into the transform coefficients of the host image after applying DWT, DFT, DCT or SVD transform, and this called the frequency domain watermarking. Because of the weakness of the spatial domain watermarks, watermarking in the frequency domain became more attractive as a result of its higher robustness against attacks compared to the spatial domain watermarking. To this aim a number of robust methods based on the SVD transform were introduced but these methods didn't offer good transparency and robustness against geometric attacks. Starting with Liu and Tan [8] an image watermarking method based on SVD, this method is robust against some attacks, and it is a non-blind method and it has a weak imperceptibility, Chandra et al. [1] also introduced a digital image watermarking method. This method is based on moderate modifying of the singular values of the host image. This method is weak against geometric attacks. Ganic et al. [9] proposed a method based on SVD in discrete wavelet transform (DWT) domain. The insertion procedure concerns the modifying of the singular values of the wavelet transformed sub-bands with the singular values of the mark. This scheme is a non-blind and the transparency of the watermarked image is weak. Makhloghi et al.[10] also proposed a scheme based on singular value decomposition in wavelet domain for copyright protection but his method lacks to robustness. Lin et al. [11] presented a full-band DWT domain image watermarking method based on SVD. This method has good robustness against common attacks but its drawback is that the original image is required in extraction procedure. Also the quality of the watermarked image is not good. The watermarking algorithms described in [1-8] are semi or non blind. Soumya Mukherjee and Arup Kumar Pal [12] proposed a robust watermarking scheme which employs both the Discrete Cosine Transform (DCT) and the Singular Value Decomposition (SVD), It starts with transformation of an original image into a transformed image using block based DCT. From each transformed block, the middle band DCT coefficients are selected to form a reduced transformed image and then the watermark is embedded into the constructed reduced transformed image after a suitable SVD operation. This method has good robustness against different attacks but it has a weak imperceptibility. In this paper, a new SVD-based method is proposed which gives the variety in creating the watermark under (1, 2, ..., n) blocks depending on the initiate coefficients and using the circulant's matrix properties to make Chandra algorithm [1] blind instead of non blind and turn it robust against different geometric and non geometric attacks. Organization of the paper is as follows: Section 2 explains the concept of SVD and Circulant matrices while Section 3 presents the proposed method. Section 4 throws light on the experimental results and a comparative analysis of our scheme and other schemes is given, whereas the summary of results and the conclusion is presented in Section 5.

## 2. SINGULAR VALUE DECOMPOSITION AND CIRCULANT MATRICES

### 2.1. Singular value decomposition SVD

Every real matrix $A$ can be decomposed into a product of three matrices :

$$A = U \times S \times V^t \tag{1}$$

where $U$ and $V$ are orthogonal matrices such that $U \times U^t = I$ and $V \times V^t = I$ where $I$ is the Identity matrix and $S$ is the diagonal matrix, $S = diag(\partial_1, \partial_2 \cdots)$ with $\partial_1 \geq \partial_2 \geq \cdots \geq 0$ The diagonal entries $\partial_i$ of $S$ are called the singular values of $A$, they are the eigenvalues of $A \times A^t$ or $A^t \times A$. The columns of $U$ are the left singular vectors of $A$, they are eigenvectors of $A \times A^t$ and the columns of $V$ called the right singular vectors of $A$ and they are eigenvectors of $A^t \times A$.

### 2.2. Circulant matrices

The circulant matrix $C = cir(c)$ associated to the vector $c \in R^n$ is the $n \times n$ matrix whose rows are given by iterations of the shift operator T acting on c, its $K^{th}$ row is $T^k c$, $k = 1,..,n$
For example if $c = (c_1, c_2, c_3, c_4)$, the $4 \times 4$ circulant matrix

$$C = cir(c) \text{ is giving by } \begin{pmatrix} c_1 & c_2 & c_3 & c_4 \\ c_4 & c_1 & c_2 & c_3 \\ c_3 & c_4 & c_1 & c_2 \\ c_2 & c_3 & c_4 & c_1 \end{pmatrix} \tag{2}$$

## 3. PROPOSED METHOD

We consider a circulant matrix $c = cir(c_1, c_2, c_3, c_4)$

The matrix $CC^t = C^tC$ is positive symmetric matrix so its spectral decomposition coincides with its SVD decomposition that is $CC^t = U_0 diag(\delta_1, \delta_2, \delta_3, \delta_4) U_0^t$ with

$$\delta_1 = (c_1 + c_2 + c_3 + c_4)^2$$
$$\delta_2 = (c_1 - c_2 + c_3 - c_4)^2 \quad (3)$$
$$\delta_3 = \delta_4 = (c_1 - c_3)^2 + (c_2 - c_4)^2$$

are the singular values and $U_0$ is the constant matrix :

$$U_0 = \begin{pmatrix} 1/2 & -1/2 & 0 & -\sqrt{2}/2 \\ 1/2 & 1/2 & -\sqrt{2}/2 & 0 \\ 1/2 & -1/2 & 0 & \sqrt{2}/2 \\ 1/2 & 1/2 & \sqrt{2}/2 & 0 \end{pmatrix} \quad (4)$$

If $A$ is an image of size $4m \times 4m$, to every vector $c = (c_1, c_2, c_3, c_4)$ is associated a $4 \times 4$ circulant matrix $C_1 = cir(c_1)$ and a watermark as $4m \times 4m$ matrix with one block

$$W_1 = \begin{pmatrix} C_1 C_1^t & 0 & . & 0 \\ 0 & 0 & & . \\ . & & . & . \\ 0 & . & . & 0 \end{pmatrix} \quad (5)$$

To obtain a watermark $W_k$ with $k$ blocks

$$W_k = \begin{pmatrix} C_1 C_1^t & 0 & . & . & . & 0 \\ 0 & C_2 C_2^t & & & & . \\ . & & . & & & . \\ . & & & C_k C_k^t & & . \\ . & & & & 0 & . \\ 0 & . & . & . & . & 0 \end{pmatrix} \quad (6)$$

### 3.1. Watermark insertion procedure

To watermark a given original image $A$ of size $4m \times 4m$ we will use a watermark with one block as following:
1. Take $c = (c_1, c_2, c_3, c_4)$ such that $\partial_4 \geq \partial_3 \geq \partial_2 \geq \partial_1$
2. Construct the watermark of size $4m \times 4m$

$$W_1 = \begin{pmatrix} C_1 C_1^t & 0 & . & 0 \\ 0 & 0 & & . \\ . & & . & . \\ 0 & . & . & 0 \end{pmatrix} \quad (7)$$

3. Apply SVD on $A$:
$$A = U \times S \times V^t \text{ with } S = diag(S_i) \quad (8)$$
4. Apply SVD on $W_1$:

$$W_1 = \begin{pmatrix} U_0 & 0 \\ 0 & I \end{pmatrix} \begin{pmatrix} \partial & 0 \\ 0 & 0 \end{pmatrix} \begin{pmatrix} U_0^t & 0 \\ 0 & I \end{pmatrix} \quad (9)$$

With $\partial = diag(\delta_1^1, \delta_2^1, \delta_3^1, \delta_4^1)$ and $I$ is $4(m-1) \times 4(m-1)$ identity matrix.

Put
$$Y_i = S_i + \alpha \partial_i' \quad (10)$$

with $\partial_1' = \delta_1^1$, $\partial_2' = \delta_2^1$, $\partial_3' = \delta_3^1$, $\partial_4' = \delta_4^1$ and $\forall i > 4\ \partial_i' = 0$

So
$$A^* = U \times diag(Y_i) \times V^t \quad (11)$$

is the watermarked image.

## 3.2. Watermarking detection and extraction procedure

We don't require the original image $A$ to detect the watermark, we only require the watermarked image $A^*$, the scaling factor $\alpha$ and the key $S_i = (S_1, S_2, S_3, S_4)$ formed by the first four values of $S$.

1. Apply SVD to $A^*$
$$A^* = U^* \times S^* \times V^{*t} \quad (12)$$
2. Calculate
$$x_i = \frac{S_i^* - S_i}{\alpha} \quad (13)$$

for the first four elements.
If $x_3 = x_4$ then the mark is detected else the watermark is not present on the image.

To extract the mark we compute:
$$W^* = \begin{pmatrix} U_0 & 0 \\ 0 & I \end{pmatrix} \begin{pmatrix} X & 0 \\ 0 & 0 \end{pmatrix} \begin{pmatrix} U_0^t & 0 \\ 0 & I \end{pmatrix} \quad (14)$$

where $X = diag(x_1, x_2, x_3, x_4)$ and $I$ is $4(m-1) \times 4(m-1)$ identity matrix.

**Remarks**:
1. If we use a watermark $W_k$ with $k$ blocks, to detect or extract the watermark we only require the scaling factor $\alpha$ and a key $K_2 = (S_1, ..., S_{4k})$ of length 4k which contains the first 4k values of S. In this case the sequence $X = (x_i)$ is of length 4k and the mark is detected if $x_{4i-1} = x_{4i}$ for $i = 1, ..., k$.

2. In Chandra algorithm [1], to extract the watermark (W), $U_w$, $V_w$ are required, in the proposed scheme $U_w = V_w$ is a constant matrix and independent of the watermark, thus our proposed algorithm is blind.

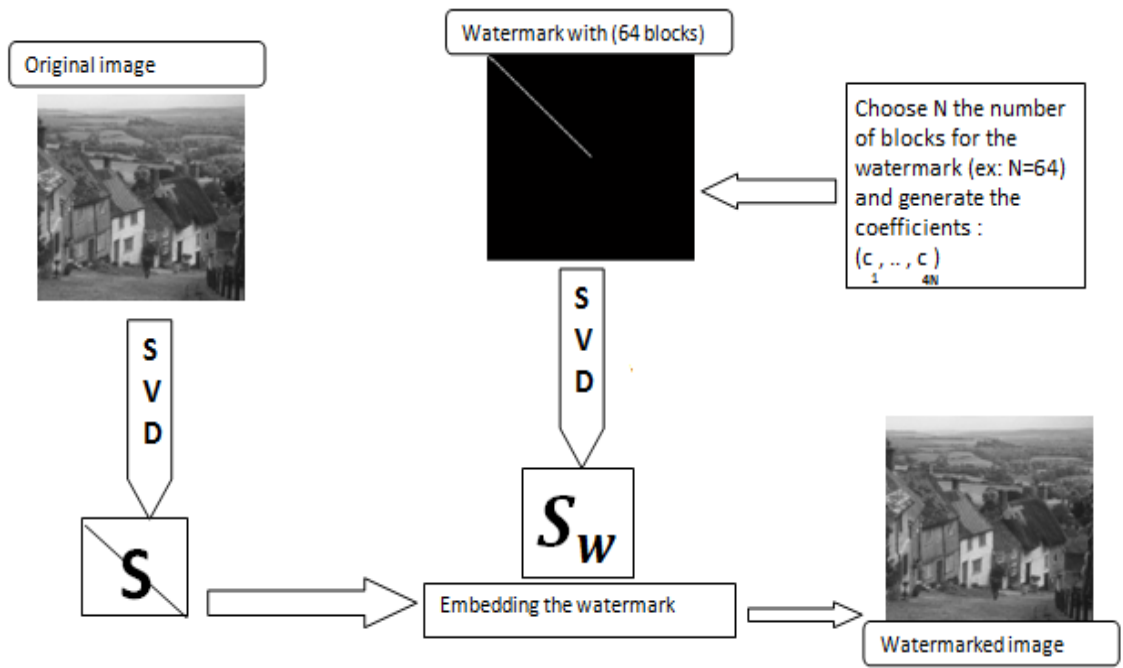

Figure 1. The Proposed watermarking scheme.

Our proposed watermarking method can be concluded in (Figure 1). As it is shown to construct the watermark we first choose N the number of blocks of the watermark then we generate the coefficients ($c_1,...,c_{4N}$) and by following the steps mentioned above we embed the watermark into the original image.

## 4. EXPERIMENTAL RESULTS

The proposed scheme is implemented using MATLAB. Six 512×512 images Lena, Goldhill, Baboon, Barbara, Peppers, Boat were used in the simulation (figure 2). The signature is a 512 × 512 image composed with N blocks in its diagonal.

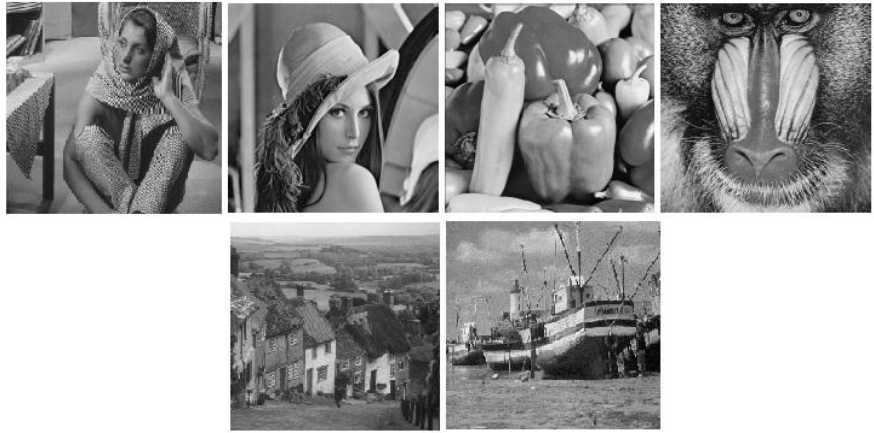

Figure 2. Original test images

To evaluate the quality of the watermarked images we use the PSNR measure defined as:

$$PSNR = 10\log_{10}(\frac{255^2}{MSE})db \qquad (15)$$

with $MSE$ is the mean squared error between the original and watermarked image. The PSNR values of the watermarked images by our method indicate that our method in general achieves very good quality as it shown in (Table 1).

Table 1. Relationship of number of blocks of the watermark and PSNR of different watermarked images. (dB) Sf sets to 0.06

| Nb blocks / Image name | 1 | 3 | 5 | 10 | 30 | 64 | 80 | 100 | 128 |
|---|---|---|---|---|---|---|---|---|---|
| Lena | 56.6994 | 55.5982 | 55.7382 | 55.0831 | 55.8089 | 56.0630 | 55.4090 | 55.4366 | 55.3805 |
| Goldhill | 57.2825 | 55.5499 | 55.4955 | 55.1460 | 54.6665 | 54.6762 | 52.9705 | 52.8710 | 52.8454 |
| Baboon | 55.6458 | 54.6902 | 55.0781 | 55.1058 | 51.6568 | 50.4369 | 49.4199 | 49.1576 | 49.2020 |
| Barbara | 56.9139 | 55.5102 | 57.1737 | 55.2064 | 53.3403 | 51.5035 | 50.2628 | 50.1065 | 50.0764 |
| Peppers | 55.9094 | 55.2634 | 55.1153 | 53.5932 | 54.5946 | 55.4747 | 56.2183 | 56.2190 | 56.2759 |
| Boat | 57.1737 | 55.6458 | 55.2064 | 54.6902 | 51.5035 | 52.9705 | 52.8710 | 52.8454 | 52.8710 |

Its clear from (Table1) that the proposed method preserves good transparency for the watermarked images.

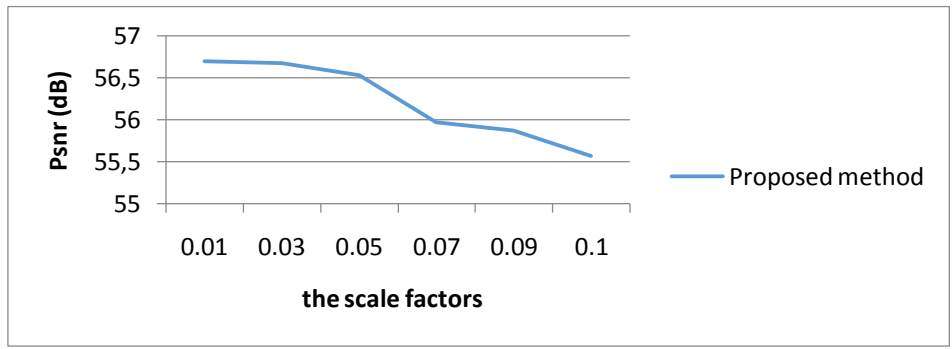

Figure 3. Relationship of the scale factors (α) and PSNR for the proposed method

(Figure 3) shows a relation between α and transparency in terms of the PSNR value. We can notice that our method has a good PNSR values and it reach its optimal value when the scale factors (α) is between 0.01 and 0.03.

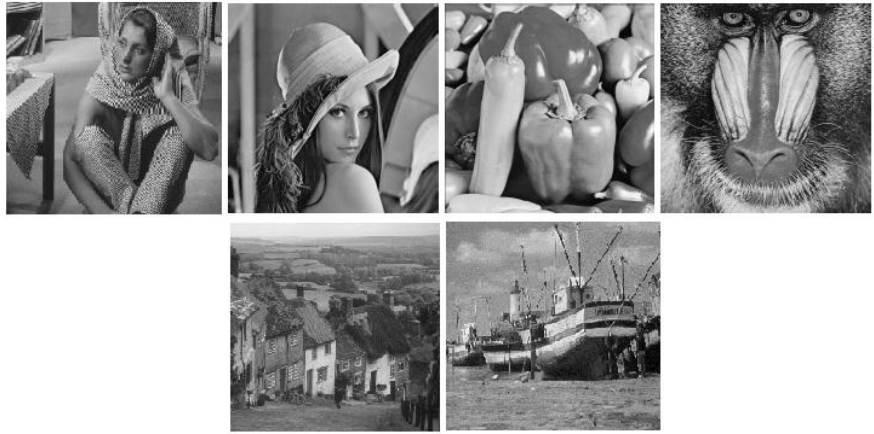

Figure 4. Watermarked images

(Figure 4) shows the watermarked images used in the simulation, it can be seen that there is no perceptual difference between original and watermarked images that is also supported by good PSNR value. It shows that the good imperceptibility is obtained by proposed technique.

We use normalized correlation (NC) to evaluate the quality of the extracted watermark, and this measure is defined by:

$$NC(W,W') = \frac{1}{W_h \times W_w} \sum_{i=0}^{W_h-1} \sum_{j=0}^{W_w-1} W(i,j) \times W'(i,j) \qquad (16)$$

with $W_h$ and $W_w$ are the height and width of the watermarked image, respectively. $W(i,j)$ and $W'(i,j)$ denote the coefficients of the inserted watermark and the extracted watermark respectively.

As we can see in (Table 2) our proposed method is robust against different geometric and non-geometric attacks, and we can notice also that the relationship between the number of blocks of the watermark and the resulted $NC$ generally are inversely proportional except for the rotation attack.

Table 2. Resulted NC for different image attacks using variation of watermarks.

| Attacks | Jpeg | | | speckle | imsharpen | Smooth | Rotation | | salt & pepper | FFT | Filtre median | translate | Gaussian filter |
|---|---|---|---|---|---|---|---|---|---|---|---|---|---|
| Nb blocks | 50 | 60 | 90 | 0.04 | | | 3° | 5° | 0.02 | | | [20 35] | hsize = [5 5] sigma = 2 |
| 1 | 0.9941 | 0.9941 | 0.9941 | 0.9985 | 0.9996 | 0.9993 | 0.9684 | 0.8477 | 0.9998 | 0.9941 | 1.0000 | 0.9989 | 0.9994 |
| 3 | 0.9577 | 0.9577 | 0.9578 | 0.9952 | 0.9982 | 0.9982 | 0.9517 | 0.8560 | 0.9997 | 0.9581 | 0.9998 | 0.9941 | 0.9974 |
| 5 | 0.9236 | 0.9236 | 0.9238 | 0.9928 | 0.9972 | 0.9975 | 0.9471 | 0.8634 | 0.9996 | 0.9246 | 0.9993 | 0.9906 | 0.9947 |
| 10 | 0.8596 | 0.8598 | 0.8603 | 0.9900 | 0.9931 | 0.9941 | 0.9454 | 0.8822 | 0.9993 | 0.8623 | 0.9962 | 0.9882 | 0.9756 |
| 30 | 0.7551 | 0.7559 | 0.7584 | 0.9885 | 0.9830 | 0.9489 | 0.9582 | 0.9195 | 0.9985 | 0.7656 | 0.9615 | 0.9889 | 0.8280 |
| 64 | 0.7235 | 0.7249 | 0.7280 | 0.9867 | 0.9796 | 0.8864 | 0.9711 | 0.9423 | 0.9978 | 0.7373 | 0.9226 | 0.9911 | 0.7058 |
| 80 | 0.7102 | 0.7123 | 0.7143 | 0.9825 | 0.9782 | 0.8459 | 0.9740 | 0.9465 | 0.9961 | 0.7243 | 0.8979 | 0.9927 | 0.6360 |
| 100 | 0.7082 | 0.7107 | 0.7121 | 0.9807 | 0.9781 | 0.8389 | 0.9755 | 0.9487 | 0.9952 | 0.7220 | 0.8939 | 0.9929 | 0.6243 |
| 128 | 0.7072 | 0.7100 | 0.7115 | 0.9803 | 0.9781 | 0.8371 | 0.9757 | 0.9491 | 0.9945 | 0.7213 | 0.8924 | 0.9928 | 0.6214 |

Table 3. Comparison of PSNR for Lai et al. [13], Tsai et al [14] and our algorithm.

| Method | the scale factors | | | | |
|---|---|---|---|---|---|
| | 0.01 | 0.03 | 0.05 | 0.07 | 0.09 |
| Lai et al .[13] | 51.14 | 51.14 | 50.89 | 49.52 | 47.49 |
| Tsai et al [14] | 47 | 37 | 33 | 28 | about 25 |
| Proposed method | 56.70 | 56.68 | 56.53 | 55.97 | 55.87 |

(Table 3) shows the comparison of PSNR for two other algorithms and our algorithm. The values of the scale factors, (SFs) are carried out with constant range from 0.01 to 0.09 with an interval of 0.02. Size of host images are 256×256 for Lai et al. [13], and 512×512 for Tsai et al. [14] and our scheme.

Table 4. The comparison of robustness and imperceptibility (dB) for our scheme and Soumya et al. [12] under various image processing attacks.

| Attack | NC of the extracted watermark (Proposed Scheme) | PSNR of the Attacked watermarked image (Proposed Scheme) | NC of the extracted watermark [12] | PSNR of the Attacked watermarked image [12] |
|---|---|---|---|---|
| Gaussian Lowpass Filtering (3×3) | **0.9991** | **56.1734 dB** | 0.9974 | 41.2799 dB |
| Average Filtering | **0.9946** | **52.6980 dB** | 0.8253 | 35.7368 dB |
| Image Noising by salt-and-pepper noise | **0.9927** | **50.1413 dB** | 0.9009 | 24.7876 dB |
| image enhancement by histogram equalization | 0.9122 | **40.0745 dB** | **0.9254** | 12.1774 dB |
| Center-cropped attack (64×64 pixels) and filled with pixel value 0 | 0.9178 | **41.4920 dB** | **0.9417** | 20.8188 dB |
| Center-cropped attack (64×64 pixels) and filled with pixel value 255 | **0.9582** | **24.7561 dB** | 0.8983 | 15.0842 dB |
| Center-cropped attack (128×128 pixels) and filled with pixel value 0 | 0.8793 | **26.3998 dB** | **0.8979** | 15.3449 dB |
| Center-cropped attack (128×128 pixels) and filled with pixel value 255 | **0.9577** | **13.8987 dB** | 0.8743 | 8.8648 dB |
| JPEG Compression (QF=25) | **0.9979** | **55.4146 dB** | 0.9281 | 35.9800 dB |
| JPEG Compression (QF=50) | 0.9979 | **55.4132 dB** | 1 | 38.7407 dB |

**The bold values indicates the best values comparing with the others.**

(Table 4) presents the comparison of robustness and imperceptibility (dB) for our scheme and Soumya et al. [12] under various image processing attacks, besides the robustness results the proposed scheme achieved high imperceptibility compared with Soumya et al. [12]

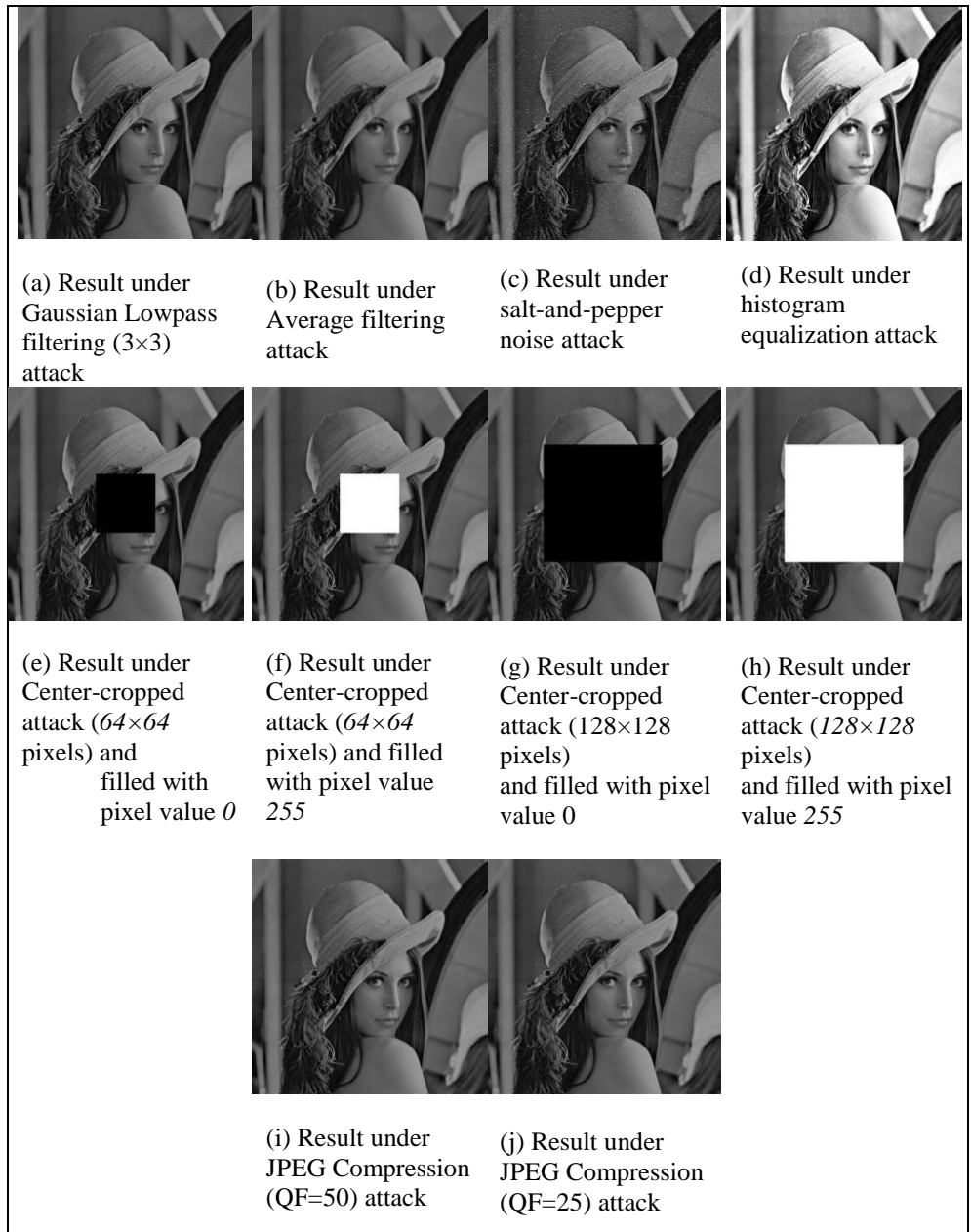

(a) Result under Gaussian Lowpass filtering (3×3) attack

(b) Result under Average filtering attack

(c) Result under salt-and-pepper noise attack

(d) Result under histogram equalization attack

(e) Result under Center-cropped attack (*64×64* pixels) and filled with pixel value *0*

(f) Result under Center-cropped attack (*64×64* pixels) and filled with pixel value *255*

(g) Result under Center-cropped attack (128×128 pixels) and filled with pixel value 0

(h) Result under Center-cropped attack (*128×128* pixels) and filled with pixel value *255*

(i) Result under JPEG Compression (QF=50) attack

(j) Result under JPEG Compression (QF=25) attack

Figure 5. The attacked watermarked image under various image processing attacks in the comparison with Soumya et al. [12]

(Figure 5) shows the attacked images of the comparison with the scheme of Soumya et al.[12].

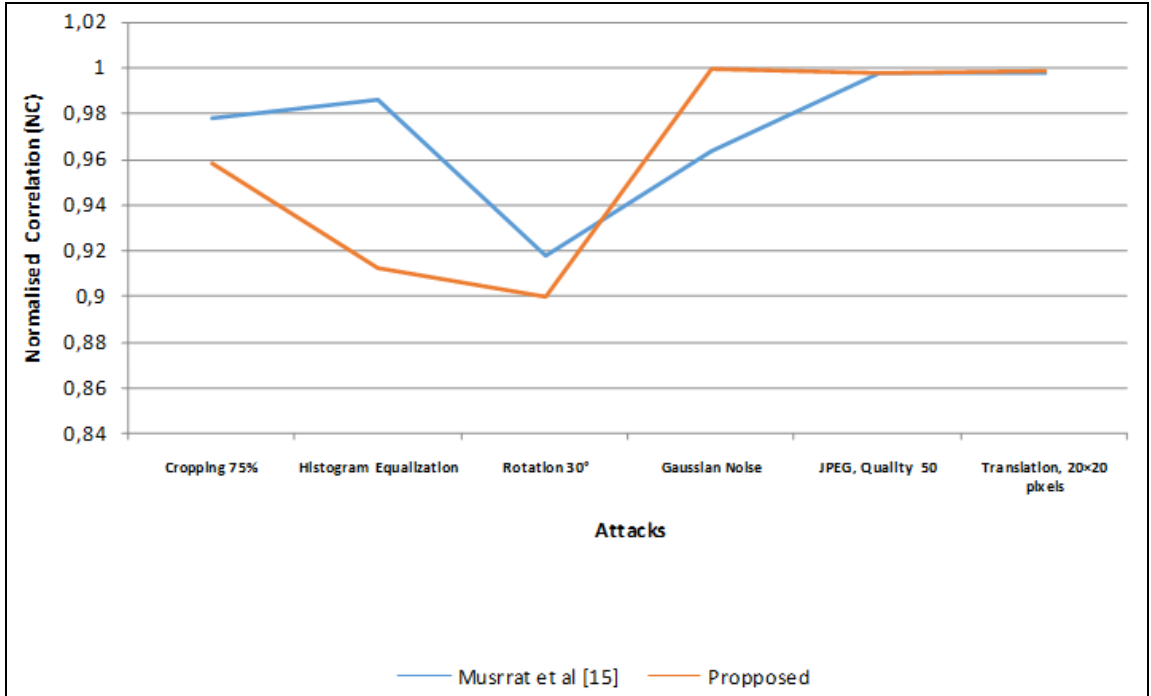

Figure 6. Comparison of our algorithm and Musrat et al [15] in term of *NC* values

(Figure 6) presents a comparison of robustness between the proposed method and Musrat et al [15] method, it shows that Musrat's scheme is more robust against such attacks as cropping 75% , histogram equalization and rotation 30° compared to our method but ours still achieves a good NC values after applying those attacks, and the minimum value of NC was 0.9 which means its robust, and for the other attacks (Gaussian noise, JPEG compression and Translation) our method perform better than Musrat's scheme.

Table 5. The comparison of robustness for our scheme, Nasrin et al [17] , Lai et al. [13] and Rastegar et al. [16].

| Attack | Proposed scheme | Nasrin et al [17] | Lai et al .[13] | Rastegar et al.[16] [a] | Rastegar et al.[16] [b] |
|---|---|---|---|---|---|
| Pepper & salt noise (0.3) | **0.9927** | 0.8926 | – | 0.7515 | 0.8258 |
| Speckle noise (var=0.01) | **0.9950** | 0.952 | – | 0.9609 | 0.9667 |
| Gaussian noise (M=0,var=0.5) | **0.9210** | 0.8935 | – | 0.7926 | 0.82 |
| Gaussian filtering (3 ×3) | **0.9990** | 0.987 | – | 0.8023 | 0.9843 |
| Median filtering (3×3) | **0.9885** | 0.982 | 0.9597 | 0.7534 | 0.9706 |
| Wiener filtering (3×3) | 0.9826 | **0.984** | – | 0.9824 | 0.9569 |
| Sharpening | **0.9966** | 0.932 | – | 0.9687 | 0.9511 |
| Histogram equalization | 0.9122 | **0.990** | 0.9862 | 0.9648 | 0.9628 |
| Gamma correction (0.7) | 0.9887 | 0.9935 | **0.9982** | – | – |
| Gamma correction (0.8) | 0.9890 | **0.9950** | – | 0.7203 | 0.9217 |
| JPEG compression Q = 30 | **0.9937** | 0.987 | – | – | – |
| JPEG compression Q = 10 | **0.9915** | 0.972 | 0.9772 | 0.9824 | 0.9843 |
| JPEG compression Q = 5 | **0.9907** | 0.952 | – | 0.8532 | 0.9354 |
| Scaling zoom(out = 0.5, in = 2) | **0.9772** | 0.948 | – | 0.5127 | 0.953 |
| Rotation (angle = 2°) | 0.9648 | **0.981** | – | 0.5068 | 0.9628 |
| Rotation (angle =−30°) | 0.8532 | **0.9823** | 0.9780 | – | – |

'–' means the attacks are not done.
The bold values indicates the best values comparing with the others.
[a] Indicates the first scheme in Rastegar et al. [16]
[b] Indicates the second scheme in Rastegar et al.[16]

(Table 5) shows the proposed results with Rastegar's schemes results and Lai's scheme results and Nasrin et al's scheme when scaling factor is 0.05. Rastegar scheme (a) represents the embedding in all sub-bands while Rastegar scheme (b) represents the embedding in LH and HL only. Our scheme performed better than Nasrin's , Lai's and Rastegar's schemes as shown in (Table 5) .

## 5. CONCLUSIONS

This paper presents a blind robust digital image watermarking scheme based on singular value decomposition and on circulant matrix for copyright protection, using the circulant matrix's properties we improved the algorithm of Chandra and turned it to a blind watermarking algorithm after it was a non blind in addition to the augmentation of its robustness. Simulation results indicate that the proposed method achieves higher robustness compared to other known watermarking methods. The proposed method is robust against a wide range of common image processing attacks.